\documentclass{article}

\usepackage{microtype}
\usepackage{graphicx}
\usepackage{subcaption}
\usepackage{booktabs} 
\usepackage{comment}
\usepackage{enumitem}
\usepackage{amsmath}
\usepackage{amsfonts}
\usepackage{xspace}

\usepackage{hyperref}

\newcommand{\showcomments}{no}
\newcommand\zhen[1]{
    \ifthenelse{\equal{\showcomments}{yes}}{\textcolor{blue}{[zhen: #1~]}}{\ignorespaces}
}
\newcommand\daiyaan[1]{
    \ifthenelse{\equal{\showcomments}{yes}}{\textcolor{red}{[daiyaan: #1~]}}{\ignorespaces}
}

\newcommand\greg[1]{
    \ifthenelse{\equal{\showcomments}{yes}}{\textcolor{green}{[greg: #1~]}}{\ignorespaces}
}

\newcommand\mason[1]{
    \ifthenelse{\equal{\showcomments}{yes}}{\textcolor{purple}{[mason: #1~]}}{\ignorespaces}
}

\newcommand\yida[1]{
    \ifthenelse{\equal{\showcomments}{yes}}{\textcolor{orange}{[yida: #1~]}}{\ignorespaces}
}

\newcommand{\sysname}{\mbox{\textsc{PipeFill}}\xspace}

\usepackage[accepted]{mlsys2024}

\mlsystitlerunning{\sysname: Using GPUs During Bubbles in Pipeline-parallel LLM Training}

\begin{document}

\twocolumn[
\mlsystitle{\sysname: Using GPUs During Bubbles in Pipeline-parallel LLM Training}



\mlsyssetsymbol{intern}{*}

\begin{mlsysauthorlist}
\mlsysauthor{Daiyaan Arfeen}{cmu,intern}
\mlsysauthor{Zhen Zhang}{aws}
\mlsysauthor{Xinwei Fu}{aws}
\mlsysauthor{Gregory R. Ganger}{cmu}
\mlsysauthor{Yida Wang}{aws}
\end{mlsysauthorlist}

\mlsysaffiliation{cmu}{Carnegie Mellon University}
\mlsysaffiliation{aws}{Amazon Web Services}


\mlsyskeywords{Machine Learning, MLSys}

\vskip 0.3in

\begin{abstract}
Training Deep Neural Networks (DNNs) with billions of parameters generally involves pipeline-parallel (PP) execution.
Unfortunately, PP model training can use GPUs inefficiently, especially at large scale, due to idle GPU time caused by \emph{pipeline bubbles}, which are often 15--30\% and can exceed 60\% of the training job's GPU allocation.
To improve the GPU utilization of PP model training, this paper describes \sysname, which fills pipeline bubbles with execution of \emph{other} pending jobs.
By leveraging bubble GPU time, \sysname{} reduces the GPU utilization sacrifice associated with scaling-up of large-model training.
To context-switch between fill jobs and the main training job with minimal overhead to the main job, and maximize fill job efficiency, \sysname{} carefully fits fill job work to measured bubble durations and GPU memory availability, introduces explicit pipeline-bubble instructions, and orchestrates placement and execution of fill jobs in pipeline bubbles.
Experiments show that \sysname can increase overall utilization by up to 63\% for GPUs used in large-scale LLM training, with $<$2\% slowdown of the training job, and 5--15\% even for low-scale LLM training.
For large-scale LLM training on 8K GPUs, the 63\% increase translates to up to 2.6K additional GPUs worth of work completed.

\daiyaan{\begin{itemize}
    \item evaluate slowdown causes and relative effect (free-memory, intermittent execution, lack of warmup), modify Executor to try to address them (e.g. non-uniform slowdown from lack of warmup across job types so use different \% bubble duration, try increase warmup by specifically targetting more batches on fewer layers)
    \item implement main job optimizer state offloading to increase free-memory
    \item implement FSDP for fill-jobs, evaluate increase in performance and max model size, augment scheduler to consider fill-job scaling
    \item improve story around scheduler, exploit heterogeneity of bubble free-memories and do ablation of scheduler heterogeneity-awareness 
    \item evaluate using better GPUs, higher PCI-e bandwidth, NVIDIA C2C
    \item fix implementation edge-cases (e.g. real bubble-characterization phase, execution of fill-jobs during all bubbles + synchronization times)
\end{itemize}}
\end{abstract}
]



\printAffiliationsAndNotice{\textsuperscript{*}Work done during internship at Amazon Web Services.}  

\section{Introduction}
\label{sec:intro}

DNN models with billions of parameters have exploded in popularity with the emergence of generative AI applications.
For example, popular large-language models (LLMs), such as GPT~\cite{gpt3} and LLaMA~\cite{touvron2023llama,touvron2023llama2}, are creating disruptive change in many domains.
But training such models can take several weeks or months even using thousands of GPUs\footnote{This paper uses ``GPU'' or ``device'' to refer to any computation accelerator for deep learning jobs, such as GPUs, TPUs~\cite{tpu}, or AWS Trainium.}.

A common approach~\cite{megatron,zheng2022alpa} of training on thousands of GPUs is to employ a combination of parallelization techniques. Pipeline-parallelism (PP)~\cite{pipedream,gpipe} is used to partition the model across multiple nodes, creating a pipeline of stages. The full pipeline is then replicated using data-parallelism, allowing for parallel processing of multiple data samples. Within each pipeline stage, tensor-parallelism is applied to partition the model weights, enabling parallel computation.
Pipeline parallelism operates by partitioning a model into pipeline stages, typically assigning one stage per node or per GPU. Each minibatch of data is further divided into smaller subsets called \textit{microbatches}. 
The forward and backward passes for each microbatch are then executed in a pipelined manner across the stages.

\begin{figure}[t]
\centering
\includegraphics[width=7.5cm,keepaspectratio]{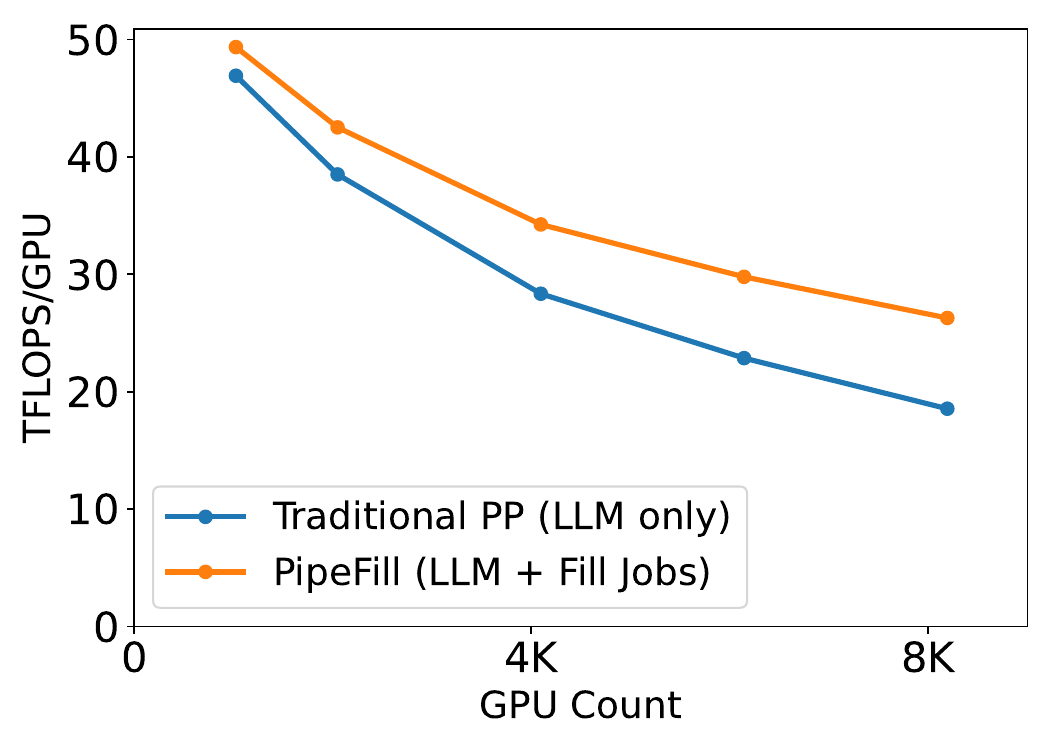}
\centering
\caption{
Utilization of LLM training GPUs.
{
\small
The lines correspond to scaling out training of a 40B-parameter LLM from 1K~GPUs to 8K~GPUs to reduce training time from 82~days (1K) to 34~days (4K) to 26~days (8K).
Traditionally, the increasing pipeline bubbles when scaling out leads to over 60\% lower GPU utilization at 8K.
\sysname{} is able to fill much of that bubble GPU time with useful work, without slowing the LLM training.
Section~\ref{sec:setup} details the experimental setup. 
}
}
\label{fig:main}
\end{figure}

Unfortunately, such highly-parallelized training can use GPUs inefficiently especially at large scale, because too much GPU time may be wasted on \emph{pipeline bubbles}.
Pipeline bubbles occur because the pipeline must be fully drained and then restarted for each minibatch, leading to idle time on each of the GPUs.
The greater the parallelization is, whether from longer pipelines (taking longer to fill and drain) or more pipeline replicas (reducing the number of microbatches per replica as the global minibatch size needs to be fixed), the greater the inefficiency becomes due to bubbles.
For example, Figure~\ref{fig:main} shows that a 40B-parameter auto-regressive-transformer LLM, parallelized over 8K~GPUs achieves
60\% lower TFLOPS-per-GPU than using just 1K~GPUs because of pipeline bubbles---but using only 1K GPUs would make LLM training take over 3$\times$ longer (26~days vs.\ 82~days; shown in Figure~\ref{fig:main_train}).
The overall consequence is a major tension between LLM training time and GPU cluster efficiency.


\sysname{} is a new GPU management system that mitigates this tension by filling large training jobs' pipeline bubbles with \emph{other} jobs, which we call \emph{fill} jobs.
The GPUs for any given pipeline stage switch to a fill job at the start of a bubble and switch back at the end of that bubble.
By doing so, \sysname recaptures otherwise wasted GPU time to accomplish pending inference and training jobs, which can enable 
scaling-up large-model training with much less sacrifice in GPU utilization.
Figure~\ref{fig:main} shows how, with bubble filling, \sysname{} mitigates the GPU utilization penalty as LLM training scales out.
At 8K~GPUs, for example, \sysname{} increases GPU utilization by over 45\% with a mix of training and inference fill jobs.
If using just less GPU memory intensive batch inference jobs, the GPU utilization increase grows to 63\% (see Figure~\ref{fig:main_res}).


Filling pipeline bubbles effectively requires overcoming a number of challenges.
First, fill job execution needs to be configured to fit within bubble constraints, including bubble length (to minimize inter-bubble context) and available GPU memory.
\sysname introduces a \emph{Pipeline Bubble Instruction} to collect bubble constraints,
and a \emph{Fill Job Execution Plan Algorithm} to partition a fill job into chunks prior to bubble filling as necessary.
Second, the right fill jobs need to be matched to the right GPUs, given that pipeline bubbles exhibits heterogeneous characteristics and users may have different optimization objectives.
\sysname uses a \emph{Fill Job Scheduler}, which accepts user-defined scheduling policies.
Our Fill Job Scheduler orchestrates the assignment of fill jobs to GPUs by synergizing the user-defined policy with the characterization of the main job's pipeline bubbles.


Experiments (real system and simulation) confirm that \sysname{} can recapture significant GPU utilization lost to pipeline bubbles, allowing huge DNNs (like LLMs) to be scaled out without much lower GPU efficiency consequences.
At each scale, aggregated TFLOPS/GPU (fill jobs plus LLM) is higher with \sysname{}, from 5--15\% at (slow) low-scale LLM training to over 63\% for scaled-out training, with $<$2\% slowdown of the LLM training.
Detailed analysis of different fill jobs options shows that, as expected, the limited memory and intermittent time available for fill job execution in bubbles reduces their efficiency differently--the data in Figure~\ref{fig:main} is for a fill job mix derived from an ML job trace, but using just bubble-efficient batch inference jobs increases the gains by $~\approx$50\%.
Additional results confirm that \sysname{}'s benefits are realized for both GPipe~\cite{gpipe} and 1F1B~\cite{pipedream} pipeline schedules, with moderate reduction (17\%) in benefits for 1F1B at low-scale and minimal difference ($<5$\%) at large-scale, and show fill job efficiency sensitivity to changes in bubble durations, available memory during bubbles, and fill-job scheduling policy.


{\bf Contributions.}
This paper makes four main contributions:

\begin{enumerate}[noitemsep,nolistsep]
  \setlength{\itemsep}{-0pt}

  \item
    It introduces the concept of filling pipeline bubbles in PP model training with execution of \emph{other} ML jobs; 
  \item
    It describes a system (\sysname) that realizes this concept and can recover idle GPU-time lost to pipeline bubbles;
  \item
    It introduces approaches for assigning fill jobs to pipeline bubbles, and for configuring fill job execution within its assigned bubble, to maximize efficiency of recovered GPU time. 
  \item 
    It experimentally shows that \sysname{} can significantly increase GPU utilization for scale-out LLM training without significantly harming LLM training efficiency.
\end{enumerate}

\section{Large Model Training Background}
\label{sec:background}

\label{sec:background:large_train}

\subsection{Distributed Training}
\label{sec:background:distributed_training}
Training Deep Neural Network (DNN) models consumes large amounts of GPU time and of on-device memory, exceeding the memory capacity of a single device for large DNNs. 
This is especially true, as the model training procedure requires memory space for multiple components, including optimizer states, model parameters, gradients, activation buffers, etc. For a 100-billion parameter model, holding the model instance requires at least 1.6TB of device memory~\cite{rajbhandari2020zero}. Example accelerator devices used to train DNNs are GPUs, TPUs, and AWS Trainium. 
The on-device memory of these devices ranges from 16GB to 80GB of HBM (high-bandwidth memory). This necessitates \textit{partitioning} the model instance, across potentially 100s of devices, in order for training to be possible. Partitioning the model instances means distributing the computation and memory footprints to multiple devices, and there are several existing methods (tensor parallelism and pipeline parallelism) for this with different tradeoffs. In addition, to further scale out the training, a training job replicates multiple model instances for computing different part of a dataset (i.e., data parallelism). We cover these techniques in the following.


\noindent \textbf{Tensor parallelism.} The tensor parallelism solution~\cite{megatron} distributes the computation onto multiple devices, and introduces communication operations to resolve the data dependencies. 
Due to the required communication operations introducing non-trivial overhead~\cite{megatron}, the tensor parallelism is typically used within a computing node. A computing node with multiple accelerators is equipped with high-bandwidth intra-node connections like NVLink. Applying tensor parallelism within a computing node makes the overhead of communication operation relatively low~\cite{zheng2022alpa}. However, using tensor parallelism within a node limits the size of a model, as the memory footprint of training a model instance must fit in the on-device memory of a single computing node.

\noindent \textbf{Pipeline parallelism.}
To further scale training across multiple computing nodes without introducing large communication overhead of tensor parallelism, the pipeline parallelism solution is used~\cite{megatron}. Pipeline parallelism partitions the DNN model across its layers. Each partition contains a set of model layers. Each partition is called \emph{a pipeline stage}. The input minibatch data to the model is split into multiple microbatches. Computations of microbatches are executed in a pipelined fashion across pipeline stages. For example, when the second pipeline stage is doing the computation of the first microbatch, the first stage can start the computation of the second microbatch. To not change the training semantics (e.g., parameter staleness, minibatch size), the pipeline parallelism has a synchronization before processing the next minibatch.

There is resource idling in the pipeline parallelism due to data dependencies and synchronizations. As noted above, to let the second pipeline stage start the computation of the first microbatch, the first pipeline stage must complete the computation of first microbatch and send the result to the second stage. And at the synchronization of the boundary of two minibatches, the other stages must wait until the slowest pipeline stage completes its work. For unidirectional, synchronous pipeline schedules (such as GPipe and 1F1B), the fraction of idling time is quantified as $(p-1) / (m + p - 1)$, where $p$ is the number of pipeline stages, and $m$ is the number of microbatches that splits from minibatch~\cite{megatron}. When scaling the training job over a cluster with thousands of devices, the fraction of idling hurts the resource utilization significantly (Section~\ref{sec:motivation:low_res_util}).


\noindent \textbf{Data parallelism.} 
The tensor parallelism and pipeline parallelism are mainly for fitting the training of a single model instance in device memory. Once the partition strategy (i.e., the number of devices used for tensor parallelism, and the number of pipeline stages of pipeline parallelism) is decided, the model instance is replicated using data parallelism to scale the training job to more devices. Each of the model instances of a training job takes a disjoint part of the training dataset for the computation. To make all model instance replicas synchronized, collective all-reduce communication across replicas is required at the end of every minibatch execution.

\vspace{-3mm}
\subsection{Combined Parallelism for LLM Training} 
\begin{figure}[t]
  \centering
  \includegraphics[width=\linewidth]{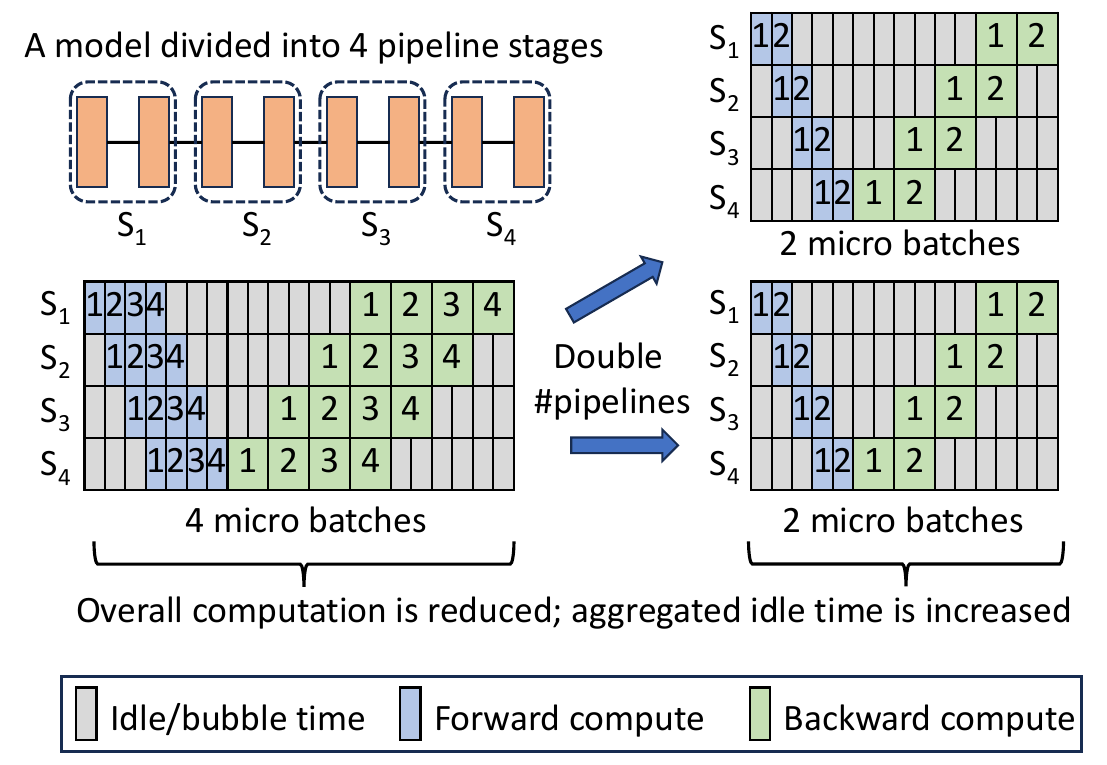}
  \caption{Pipeline parallelism combined with data parallelism. Replicating the pipeline (double the number of GPUs) with the overall minibatch size fixed (at 4~microbatches) leads to shorter per-minibatch execution time but also a larger fraction of GPU time lost to pipeline bubbles.
  }
  \label{fig:motivation_pipeline_bubbles}
\end{figure}

\daiyaan{we can try to cut this too}
For training large DNNs, such as LLMs, it is common to combine all three parallelization techniques. Figure~\ref{fig:motivation_pipeline_bubbles} illustrates an 8-layer model (upper left) partitioned into 4~pipeline stages for execution.
The per-minibatch execution timeline is illustrated below the model, when there are four microbatches, with time flowing left-to-right and each rectangle indicating what the compute node for a given stage is doing during that time.
The light-gray pipeline bubbles occur when a stage is waiting for input from other pipeline stages in order to utilize the assigned GPUs.

To scale up with a fixed total computation workload per model updating, users usually augment the data parallelism degree while reducing the number of microbatches per pipeline replica, to ensure their product remains constant.
The right side of Figure~\ref{fig:motivation_pipeline_bubbles} gives an example of doubling the number of compute nodes using data parallelism.
%
As shown in the figure, the number of the microbatches for each pipeline to process is reduced from 4 to 2 (half for each). 
There are also more advanced methods like automatically explore the best possible sharding strategies~\cite{zheng2022alpa} which may achieve better training throughput than manual decisions.
However, all of them are inherently bound by the total computation workload per model updating, hence getting diminishing return while increasing the compute resource.
This paper introduces an orthogonal way to recollect the waste GPU resource at large-scale training by introducing independent fill jobs.

As discussed earlier, the fraction of execution time of each model update spent in pipeline bubbles is $(p-1) / (m+p-1)$, where $p$ is the number of pipeline stages and $m$ is the number of microbatches. 
In the figure, the bubble fraction increases by about 40\%.






\section{Motivation and Our Solution}
\label{sec:motivation}

\subsection{Pipeline Bubbles Lower GPU Utilization}
\label{sec:motivation:low_res_util}

In pipeline parallelism, pipeline bubbles arise from data dependencies and synchronizations, resulting in inefficient GPU utilization during periods of resource idleness.
The fraction of idle GPU time of each model update spent in pipeline bubbles is $(p-1) / (m+p-1)$, where $p$ is the number of pipeline stages and $m$ is the number of microbatches. 
The issue of inefficient GPU utilization due to pipeline bubbles becomes particularly apparent when scaling up the training of larger models across more computing nodes. 
This is because the scale-up usually leads to a higher number of pipeline groups and a decrease in the number of microbatches for each pipeline group.

When applying combined parallelism, the training of larger models necessitates a larger number of pipeline stages.
Since tensor parallelism is applied within each computing node to mitigate communication overhead, the maximum degree of tensor parallelism is constrained by the number of GPUs within a single computing node. 
Once the tensor parallelism degree reaches its maximum, users can only increase the number of pipeline stages to partition larger models, ensuring that each partition fits within the GPU memory capacity.

Scaling up training across a larger number of computing nodes results in a reduction in the number of microbatches.
Once the degrees of tensor and pipeline parallelism are determined, users augment the data parallelism degree to distribute training across more GPUs, thereby enhancing training throughput.
However, when training large language models (LLMs), the total computation workload for each round of model updating is usually set by machine learning experts and remains constant, regardless of the size of the computing cluster.
For example, both LLaMA-1~\cite{touvron2023llama} and LLaMA-2~\cite{touvron2023llama2} training use 4 million tokens for each model update.
People are reluctant to increase the number of tokens for each model update, as it can hurt model quality at the end of training~\cite{mccandlish2018empirical}.
With a fixed total computation workload for each round of model updating, increasing the data parallelism degree results in a smaller number of microbatches.


In summary, inefficient GPU utilization caused by pipeline bubbles is inevitable when employing pipeline parallelism, and it becomes particularly noticeable when scaling up the training of large models.

\subsection{Solution: Fill Bubbles w/ Independent Jobs}
\label{subsec:challenges}

How can idle GPU time resulting from pipeline bubbles be utilized to improve GPU utilization? 
Existing works fill dependent jobs of the training job running with pipeline parallelism into the pipeline bubbles.
For instance, PipeFisher~\cite{pipefisher} accelerates convergence by utilizing the pipeline bubbles to execute K-FAC, a second-order optimization method based on the Fisher information matrix.
Similarly, Bamboo~\cite{bamboo} enhances training resilience at a minimal cost by filling redundant computations, where one node performs computations not only on its own layers but also on some layers in its neighbor, into the pipeline bubbles.
However, the jobs filled into the pipeline bubbles by existing works are dependent on the training job running with pipeline parallelism providing extra work (second-order computation in the case of PipeFischer, redundant computation in the case of Bamboo).
As a result, tailored scheduling for each fill job is needed to avoid performance penalties. Additionally, these prior works are only applicable to specific types of training jobs (training jobs optimized using K-FAC in the case of PipeFischer, jobs running on faulty/spot machines in the case of Bamboo).
Fundamentally, pipeline bubbles exist due to data dependencies within the computation of pipeline parallelism.
Our key insight is that, rather than directly addressing the data dependency issue within a training job pattern or introducing other dependencies by filling dependent jobs, we leverage independent jobs, unrelated to the training job running with pipeline parallelism, to fill the pipeline bubbles.
Specifically, we remove the constraint that the training job must execute exclusively on the GPUs during the entirety of the job.
One can context-switch to a different job during the bubbles to reduce the amount of idle time of GPUs, and context-switch back to the main training job in time for the training job to experience no overhead from sharing the GPU during the pipeline bubbles.

To fill independent jobs into pipeline bubbles, we need to address the following challenges:
\begin{itemize}[leftmargin=*]

\item
\textbf{Memory Management.}
How can one fill independent jobs into the pipeline bubbles when the GPU memory is primarily occupied by the main training job?
Even during pipeline bubbles, the main training job dominates the GPU memory.
Naively filling independent jobs into pipeline bubbles without careful memory management may result in GPU OOM errors or sub-optimal performance of fill jobs.
Effective memory management is crucial not only to mitigate OOM risks but also to optimize available memory for fill jobs.
%

\item
\textbf{Context Switching.}
How can one ensure that filling independent jobs into the pipeline bubbles does not incur performance penalties for the main training job?
To maintain the performance of the main training job, only pipeline bubbles can be utilized for running fill jobs.
However, it's not guaranteed that a fill job can be completed within one bubble.
Therefore, filling independent jobs into pipeline bubbles without carefully crafted context switching may introduce performance penalties to the main training job.
%

\item
\textbf{Fill Job Scheduling.}
When faced with numerous pipeline bubbles exhibiting heterogeneous characteristics, how can one effectively schedule the filling process to align with user-specific objectives?
Pipeline bubbles across various pipeline stages or employing different scheduling algorithms exhibit distinct characteristics, such as duration and HBM availability.
Additionally, users may harbor unique optimization goals; for instance, some prioritize GPU utilization, while others emphasize meeting job deadlines promptly.
Naively scheduling the filling process without accounting for bubble characteristics and users' optimization objectives risks compromising the performance of fill jobs and falling short of users' expectations.
%

\end{itemize}


Several existing works, such as Muri~\cite{muri} and Antman~\cite{antman}, explore interleaving multiple jobs on shared GPUs.
However, these works do not specifically address scheduling alongside a main job running with pipeline parallelism, thus failing to leverage the unique characteristics for optimization.
Muri only considers job duration of each job as a constraint and assumes all jobs fit together in GPU memory.
Thus, Muri lacks support for guaranteeing the main job performance and also falls short in memory management and fill job scheduling.
Antman utilizes device statistics to assign memory caps to jobs based on priority and fills idle GPU cycles with opportunistic kernels.
However, when a main job is running with pipeline parallelism, pipeline bubbles often appear as long-running communication kernels, causing Antman to struggle in determining context switches between the main job and fill jobs.
Moreover, the main job typically consumes the majority of the memory, making simply setting memory caps insufficient for memory management.

To address the challenges of memory management and context switch, \sysname introduces a \emph{Pipeline Bubble Instruction} and a \emph{Fill Job Execution Plan Algorithm}.
The Pipeline Bubble Instruction serves to pinpoint the start and end of a bubble, while also capturing the information about available memory during its duration. 
The Fill Job Execution Plan Algorithm then utilizes this information to determine the feasibility and methodology of partitioning a fill job into chunks prior to bubble filling. 
Additionally, the algorithm determines when to offload the memory of the main job to free up space for the fill job when necessary.
To address the challenge of fill job scheduling, \sysname leverages a \emph{Fill Job Scheduler}, which accepts user-defined scheduling policies.
This feature grants users the flexibility to craft policies aligned with their optimization goals.
Our Fill Job Scheduler orchestrates the assignment of fill jobs to GPUs by synergizing the user-defined policy with the characterization of the main job's pipeline bubbles.

\section{\sysname Design and Implementation}
\label{sec:design_new}

\begin{figure}[t]
  \centering
  \includegraphics[width=\linewidth]{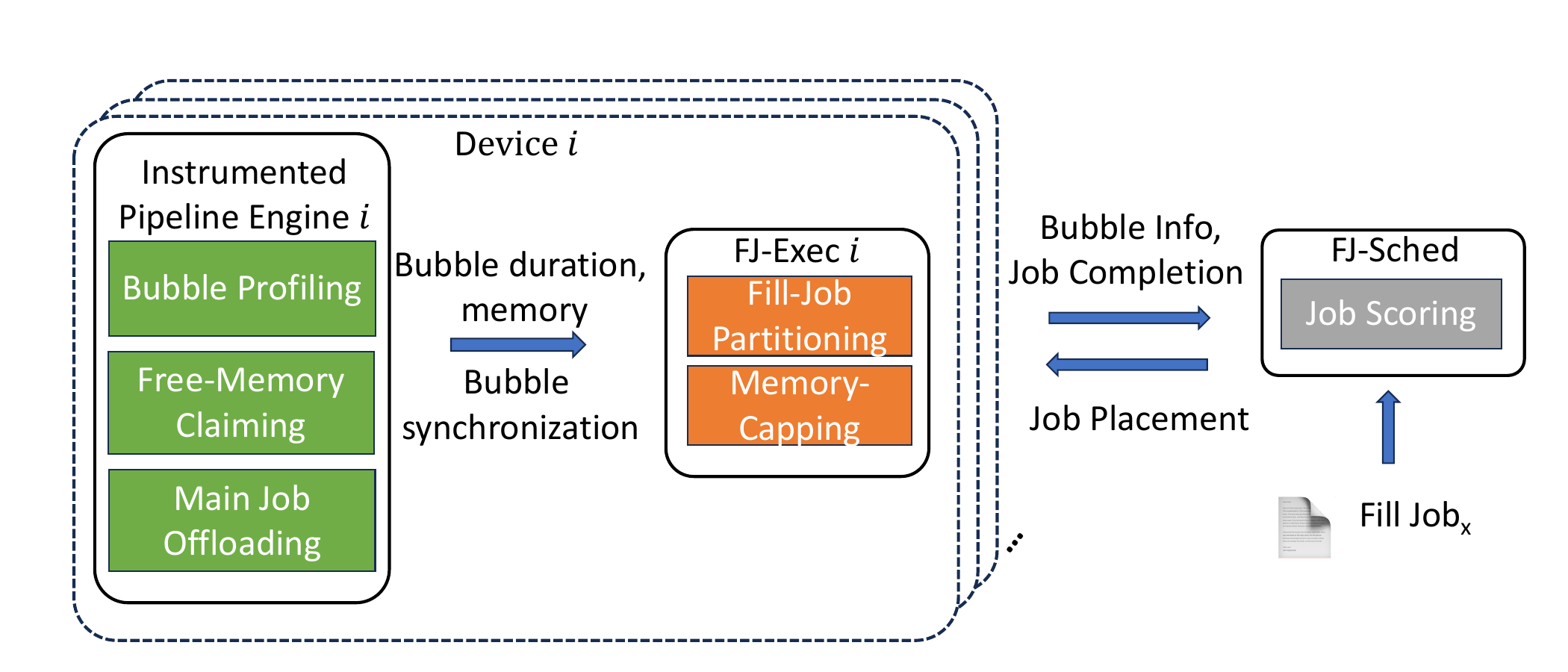}
  
  \vspace{4pt}
  \scriptsize
  \vspace{-4pt}
  
  \caption{System overview}
  \label{fig:system_overview}
\end{figure}

\subsection{\sysname Overview}


As shown in Figure~\ref{fig:system_overview}, \sysname consists of three major components: \emph{Instrumented Pipeline Engine}, \emph{Fill Job Executor}, and \emph{Fill Job Scheduler}.
The Instrumented Pipeline Engine uses the \emph{pipeline bubble instruction} to measure when a pipeline bubble begins and ends, and the available memory during a pipeline bubble.
The Fill Job Executor then leverage those information to decide the strategy of filling a job into bubbles, including whether and how to partition a filled job into chunks and whether to offload the memory of the main job to free up space for the filled job.
The Fill Job Scheduler accepts user-defined scheduling policies and schedules filled jobs onto device pipeline-bubbles to optimize the chosen policy.

\noindent \textbf{Putting together.}
Fill jobs are initially received by the Scheduler, which makes scheduling decisions about which device's pipeline bubbles to execute a fill-job on.
Scheduling policies can be defined to modify the behavior of the Scheduler.
Each device has a fill-job Executor process. Once a fill-job arrives at a device, the Executor uses profiling data to construct an execution plan for the job. The plan maximizes the throughput of the job by choosing a batch size and creating partitions of the job's computational graph that maximize the amount of work completed during the pipeline bubbles without violating bubble duration or free-memory constraints. The Executor uses synchronization primitives to decide when to begin execution of the next fill-job graph partition. 
The pipeline engine runs on every device worker. The pipeline engine uses the pipeline bubble instruction to know when a pipeline bubble is beginning so it can signal the Executor (using the aforementioned synchronization primitives) to begin execution. Before the pipeline engine signals the Executor, it tells the device memory allocator to release all transient/unused memory buffers to increase the free-memory available to the Executor and waits for any main-job offloading operations to complete.

\noindent \textbf{Fill Jobs.}
In this work, we use deep learning training and batch-inference jobs as fill jobs.
Deep learning jobs can be classified as training or inference. Training is typically not latency sensitive and is often long-running. Inference can be broken down into real-time/online inference and batch/offline inference; the former is latency sensitive, with SLOs being on the order of milliseconds, while the later is often not latency sensitive. Batch inference is not uncommon, often being used for content recommendations, data analytics, and other back-end services. Due to the intermittent property of pipeline bubbles, latency-sensitive jobs are not suitable for use as fill-jobs. Therefore, \sysname supports training and batch-inference jobs as fill-jobs. \sysname takes as input the model used for the fill-job, as well as valid batch-sizes; given the job configuration, it will attempt to execute the fill-job with maximum throughput.

\subsection{Pipeline Engine Instrumentation}
In order to not impact the main job, we must keep the context switching and execution of fill jobs completely within the duration of the pipeline bubbles. We must also know exactly how much GPU memory the fill job can use during its execution so that the main job does not experience an OOM error. In order to achieve this, \sysname augments existing pipeline engines with a new pipeline bubble instruction. Existing pipeline engines execute a sequence of pipeline instructions. These instructions include sending/receiving activations and gradients, executing forward/backward computations on specific microbatches, and synchronizing parameters. Taken together as a periodically repeating sequence, these instructions constitute a pipeline schedule, which can have multiple pipeline bubbles which appear as instructions that wait on some event (e.g. activation data to arrive from the previous stage). \sysname's bubble instruction is inserted into the schedule to indicate where large bubbles are expected to occur. \par
\noindent \textbf{Bubble characterization.} Before doing any bubble filling, the pipeline engine must determine the time duration of each pipeline bubble in the pipeline schedule and how much memory is available for the fill jobs to use. To this end, at the beginning of the main training job the pipeline engine does profiling. For each bubble instruction, the pipeline engine will wait certain amount of time (e.g. 100 ms) before proceeding to execute the next instruction. It will then observe the main job's throughput, if it is unaffected then on the next minibatch iteration it will wait 2$\times$ amount of time. This will continue until the pipeline engine observes a drop in  the main job's throughput, at which point it will know the duration of the pipeline bubble.
To profile the amount of memory available for the engine to use for the fill job during a pipeline bubble, the engine relies on PyTorch's \verb|torch.cuda.memory_allocated()| function to know how much memory is held by the main training job during the bubble; the remaining device memory is considered free, but to ensure there are no out-of-memory errors \sysname may opt only to allocate some fraction of the free memory. Additionally, to ensure transient/temporary memory buffers are not counted as allocated by the main job (and instead can be used by the fill-jobs), the engine will tell the memory allocator to free all such buffers (by calling torch.cuda.empty\_cache()). The bubble duration and free-memory capacity is later passed to the Executor so it can avoid violating those constraints. \newline
\noindent \textbf{Bubble signaling.} Once the engine has characterized the pipeline bubbles they can start to be filled. The engine starts a new Executor process (with a shared synchronization primitive) and passes the bubble information to it. When a new fill-job is sent by the Scheduler, the engine passes the job description (as well as the necessary profiles) to the Executor. Every time the pipeline engine reaches a bubble instruction, it  1) tells the memory allocator to free all unused memory 2) waits for any main job offloading operations to complete 3) signals the Executor to begin running its fill-job. \newline
\noindent \textbf{Main job offloading.} In some cases, it may be beneficial to increase the amount of free-memory available to the fill-jobs. To achieve this, \sysname enables offloading of main job data from the device to the CPU memory. In order to do this in a way that is transparent to the main job and does not sacrifice its performance, which data is offloaded must be carefully chosen and the offloading and onloading of the data must be coordinated so that the main job is never blocked on these operations. \sysname enables offloading of the main job optimizer states (e.g. moment estimates for Adam\cite{adam}) because this data is only required by the main job during the optimizer updates. The offloading of the main job data is overlapped with the forward-pass execution, and the onloading is overlapped with the gradient-synchronization; a significant amount of data can be offloaded in this fashion with no impact to the main job. The pipeline engine forward-pass and gradient-synchronization instructions are augmented to launch these operations on a separate CUDA stream.

\subsection{Executor}
The Executor is a process that executes a fill-job on a device's pipeline bubbles with maximum throughput without violating the bubble duration or free-memory constraints, ensuring that the fill-job execution has no impact on the main job performance. It does this by creating an execution plan for the fill job that chooses a batch size and partitions the job's computational graph, and it relies on signals from the pipeline engine to know when to execute the graph partitions. \par
\noindent \textbf{Execution plan.}
When created, the Executor is passed a sequence of bubble durations and free-memory capacities from the pipeline engine. This sequence describes the pipeline bubbles, the resources and durations that they each make available as well as their order. This sequence of bubbles is a cycle of bubbles that repeats every minibatch iteration of the main job. When a fill-job is passed to the Executor, it is accompanied with a set of profiles. Each profile contains the execution time and memory requirement of each node in the computational graph under a specific configuration. Configurations can be different batch sizes and different execution techniques (e.g. CPU-offloading or NVMe-offloading of parameters/gradients/optimizer states, activation checkpointing/offloading). The Executor linearizes the computational graph and its profiles, turning it into a sequence of nodes with sequential dependency. For each configuration, the Executor packs the computational graph into as few bubble cycles as possible (without violating duration and free-memory constraints). As shown in Algorithm~\ref{algm:fill}, the Executor runs a greedy algorithm that does the following: 1) replicate the graph enough times (each replica represents an iteration) that the total execution time is as high as possible without exceeding the total bubble time (lines 1-5), 2) iteratively packs as many source nodes of the remainder of the computational graph as possible into the next bubble (lines 9-15) without exceeding its duration or memory limits (line 10). This sequence of computational graph partitions represents the Executor's plan for the fill-job. 

\noindent \textbf{Bubble synchronization and memory capping.} When executing the fill-job plan, the Executor waits for signals from the pipeline engine to know when the main job has entered a pipeline bubble. When it receives a signal, it first sets a cap on the amount of device memory that it can use (by using PyTorch's \verb|cuda.set_per_process_memory_fraction| function) to the amount of free-memory available in the bubble; if the Executor somehow exceeds this memory capacity, it will experience an OOM error, but this error will be isolated to the Executor process and will not affect the main job. The Executor will execute the current graph partition on the current bubble, and then wait for the next signal from the pipeline engine.


\begin{algorithm}
\caption{Partition fill job onto bubbles}
\label{algm:fill}
\begin{algorithmic}[1] 
\STATE \textbf{Input:} A list $B$ of the bubble durations, a list $M$ of bubble free-memory capacities, a list $F$ of the graph-node durations and memory requirements
\STATE \textbf{Output:} List $P$ of graph partitions where duration of $P[i] \leq B[i \mod \text{len}(B)]$ and memory of $P[i] \leq M[i \mod \text{len}(M)]$
\STATE $F' \leftarrow F$
\WHILE{$dur(F') + dur(F) < \sum B$}
    \STATE $F' \leftarrow F' + F$
\ENDWHILE
\STATE $F \leftarrow F'$
\STATE $P \leftarrow []$
\STATE $i \leftarrow 0$
\WHILE{$\text{len}(F) > 0$}
    \STATE $P' \leftarrow []$
    \WHILE{$\text{len}(F) > 0$ \textbf{and} $dur(P') + dur(F[0]) < B[i]$ \textbf{and} mem($F[0]$) $\leq M[i]$}
        \STATE $P' \leftarrow P' + F[0]$
        \STATE $F \leftarrow F[1:]$
    \ENDWHILE
    \STATE $P \leftarrow P + P'$
    \STATE $i \leftarrow (i + 1) \mod \text{len}(B)$
\ENDWHILE
\RETURN $P$
\end{algorithmic}
\end{algorithm}

\subsection{Scheduler}
The Scheduler is the interface between the pipeline bubbles of the main job and any outside higher-level cluster schedulers, making the bubbles available as additional resources. The Scheduler is also responsible for scheduling the fill-jobs onto the pipeline bubbles. The Scheduler has access to the fill-job profiles, partitioning algorithm, and bubble descriptions of every device. Using this information, the Scheduler is able to precisely calculate any fill-job's throughput/processing-time on any device. The Scheduler exposes the scheduling policy by defining a function that takes as input a job's information (arrival time, processing-time on every possible device, and deadline) as well as the current state of all the Executors in the system, and outputs a score. When a device completes a fill-job, the Scheduler chooses which job to submit to the device by choosing the job which maximizes the score. This allows specifying a variety of different scheduling policies. For example, to specify a Shortest-Job-First policy the function can be defined as:
$$f(j, s, i) = \frac{1}{min(j.proc\_times)}$$
where $j.proc\_times$ is a list containing the job's processing times on all devices, s is the current state of all Executors, and i is the index of the Executor which is to be filled. A more complex example is a policy that minimizes the makespan, which can be specified with the function:
$$f(j, s, i) = \frac{1}{max(j.proc\_times[i], s.rem\_times)}$$
where $s.rem\_times$ is a list containing the remaining amount of time each Executor will be busy. This policy will minimize the maximum busy time across all Executors, thereby minimizing makespan. By defining the policy using weighted compositions of multiple functions, hierarchical policies can be defined that behave differently under different circumstances. For example, policies can be defined that prioritize proximity-to-deadline as a feature, but default to more standard policies (e.g. SJF, FIFO) when there are no jobs with deadlines. \par
Since the Scheduler knows how long the currently executing fill-jobs will take to complete, as well as the order in which the queued fill-jobs will be executed, users can query the Scheduler to know when a currently submitted fill-job is expected to complete or whether a fill-job's deadline can be met under current conditions. This can be used by a higher-level scheduler, which manages other resources in addition to the pipeline bubbles, to make scheduling decisions about which of its jobs can be submitted to the Scheduler.
\subsection{Implementation}
\label{sec:imp}
Our implementation is based on DeepSpeed. We augment the DeepSpeed pipeline engine with the instrumentation for bubble filling and main job offloading. The Executor is implemented as a python process, and it also uses DeepSpeed to execute the fill jobs. To support large-model fill-jobs with limited GPU free-memory, the Executor is enabled with fill-job configurations that use CPU-offloading and activation checkpointing. In particular, the Executor will consider using ZeRO-Offload\cite{zero-offload} and ZeRO-Infinity\cite{zero-infinity} to offload optimizer states, gradients, activations, and parameters of the fill-job. \par
\noindent \textbf{Main job pipeline schedule.} We consider GPipe\cite{gpipe} and 1F1B\cite{pipedream} schedules for the main job. Both schedules exhibit two-phase bubble behavior: one bubble occurs between the drain of the previous minibatch iteration and the fill of the next iteration (fill-drain), and the other bubble occurs between the forward-pass pipeline saturation and the backward pass (fwd-bwd). The fill-drain bubble of both schedules is the same, but the fwd-bwd bubbles can be different. For GPipe, the fwd-bwd bubble duration is $(num\_stages - stage\_id - 1) * (t_{fwd} + t_{bwd})$ whereas for 1F1B its duration is $(num\_stages - stage\_id - 1) * t_{bwd} + max(0, num\_stages - stage\_id - m) * t_{fwd}$. 1F1B additionally has some non-contiguous bubbles (which \sysname does not fill), which makes the total bubble time the same for both schedules.

\begin{table}[t]
\centering
\resizebox{.41\textwidth}{!}{
\begin{tabular}{ |c|c|c|c| } 
 \hline
 \textbf{size} & \textbf{model} & \textbf{\# parameters} &\textbf{job type}\\ 
 \hline
 S & EfficientNet\cite{tan2019efficientnet} & 117M & CV \\ 
 \hline
 S & Bert-base\cite{bert} & 109M & NLP \\ 
 \hline
 M & Bert-large\cite{bert} & 334M & NLP \\ 
 \hline
 M & Swin-large\cite{swin} & 779M & CV \\
 \hline
 L & XLM-Roberta-XL\cite{xlm} & 2.8B & NLP \\
 \hline
\end{tabular}
}

\vspace{4pt}
\scriptsize
  \textbf{S}: small \quad
  \textbf{M}: medium \quad
  \textbf{L}: large
  
\caption{
Fill job category.
}
\label{table:fill_jobs}
\end{table}

\section{Experimental Setup}
\label{sec:setup}

\subsection{Hardware and Simulator}

In our experiments, we use a cluster of 16 AWS EC2 \emph{p3.16xlarge} instances to run small-scale experiments and to collect traces for large-scale simulation experiments.
Each \emph{p3.16xlarge} instance contains 8 NVIDIA Tesla V100 GPUs, each of which is equipped with 16GB HBM and has 125 TFLOPS of peak compute.
GPUs on the same machine are connected in a hybrid cube-mesh topology with NVLink 2.0 300GBps interconnects, and separate machines are connected with 25 Gbps network bandwidth.

In order to evaluate our system on multiple large-scale settings, we create an event-driven simulator.
Deep learning jobs have repetitive patterns, so an accurate simulator only needs to profile a pattern once and can simulate the time and resources it takes to repeat that pattern.
Our simulator relies on profiles of the main training jobs' pipeline instructions and the fill jobs' layers (under different configurations). The events in our simulator are the arrivals and completions of fill-jobs (since these are when the state of the system can change), and we simulate the time in between these events using the profiled execution times and the job arrivals from the trace.
We validate the accuracy of the simulator against the physical cluster experiments. 

\subsection{Main Jobs}
Our physical cluster experiments use a 5B parameter LLM training job as the main job, and are executed on 16 GPUs on separate machines; this main job uses 16-stage pipeline-parallelism (no tensor-parallelism).
We also collect profiles of a 40B parameter LLM training job executed using 8-way tensor-parallelism (8 GPUs per machine) and 16-stage pipeline-parallelism (16 machines).
The simulator main job has almost the same settings as the physical cluster job, only scaled up using tensor-parallelism; consequently the bubble sizes are almost identical.
We use the profiles of the 40B model training job to seed our simulator, which we use for sensitivity studies done in simulation.

Both main jobs use sequence length of 2048 tokens per sample, and use the Adam\cite{adam} optimizer.
Both jobs also use a microbatch size of 2 and a total minibatch size of 1024 (across all data-parallel replicas). We use the GPipe schedule by default, unless otherwise specified.
Data-parallel execution has been shown to be predictable\cite{pytorch-dist}, so we run only one data-parallel replica across all our experiments, varying the number of microbatches according to different data-parallel configurations.

\subsection{Fill Jobs}
We create our fill-job traces in two steps. First, we construct a fill-job model distribution. To do this, we extract all model sizes and model types from the HuggingFace Model Hub\cite{huggingface}; we filter for models uploaded in the last year with over 100K downloads. We find that among these models, 71\% of them have less than 3B parameters, so we filter out all models that have greater than 3B parameters. Among the remaining models, we find that 10.4\% are CNNs (with the remainder being transformer models). We choose a representative set of models shown in Table \ref{table:fill_jobs}, and we set sampling probabilities to each model to match the distribution of model sizes and types from the HuggingFace Model Hub. \par 
For sampling job arrivals, we use public traces from Alibaba\cite{weng2023beware} collected on real GPU clusters. These traces provide arrival times, GPU quantities requested, service times, and quality-of-service for each job. We filter out all jobs that have "latency-sensitive" quality-of-service, and we convert GPU quantity requested and service time to GPU-hours (by multiplying the two). We filter out jobs that are greater than 9 GPU-minutes for the physical cluster experiments (leaving 55\% of all jobs) and 1 GPU-hour for the simulation experiments (leaving 81.6\% of all jobs), and we bucket the remaining GPU-hours distribution according to the sampling probabilities of the models from Table \ref{table:fill_jobs} so that every job arrival in the trace is mapped to a specific model. For smaller models (<700M parameters) we set the job to training or batch-inference with equal probability; for larger models we always set the job to batch-inference. To determine how many samples a job should process, we divide the job-size (in GPU-hours) by the max throughput that the job-type can achieve when executed in isolation on one GPU. This yields a trace that contains job arrivals, job models, job category (training vs batch-inference), and job samples.

\section{Evaluation}
\label{sec:eval_new}

\begin{figure*}[t!]
  \centering
  \begin{subfigure}{0.3\linewidth}
    \centering
    \includegraphics[width=\columnwidth]{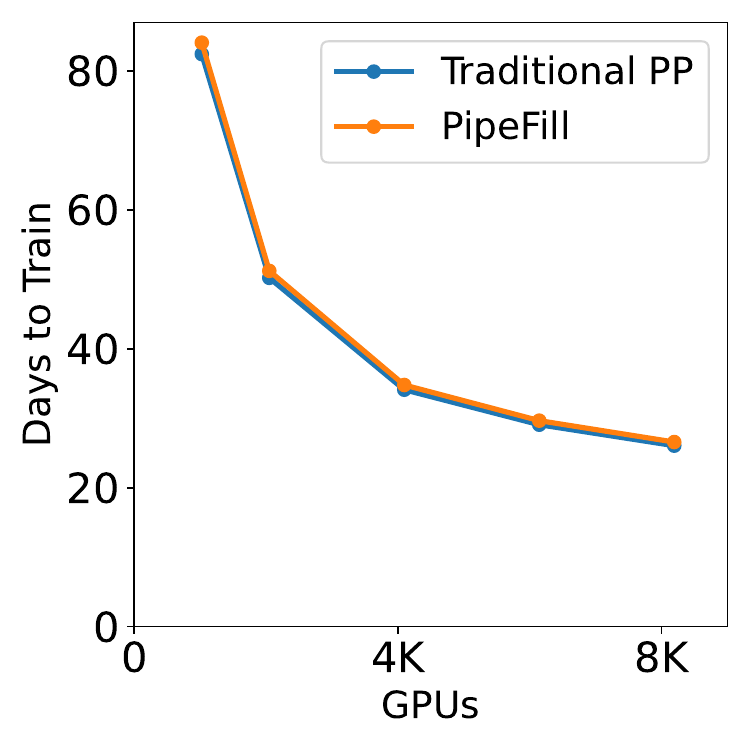}
    \caption{Main Job Training v.s. \#~GPUs}
    \label{fig:main_train}
    \vspace{-10pt}
  \end{subfigure}
  \hspace{2mm}
  \begin{subfigure}{0.31\linewidth}
    \centering
    \includegraphics[width=\columnwidth]{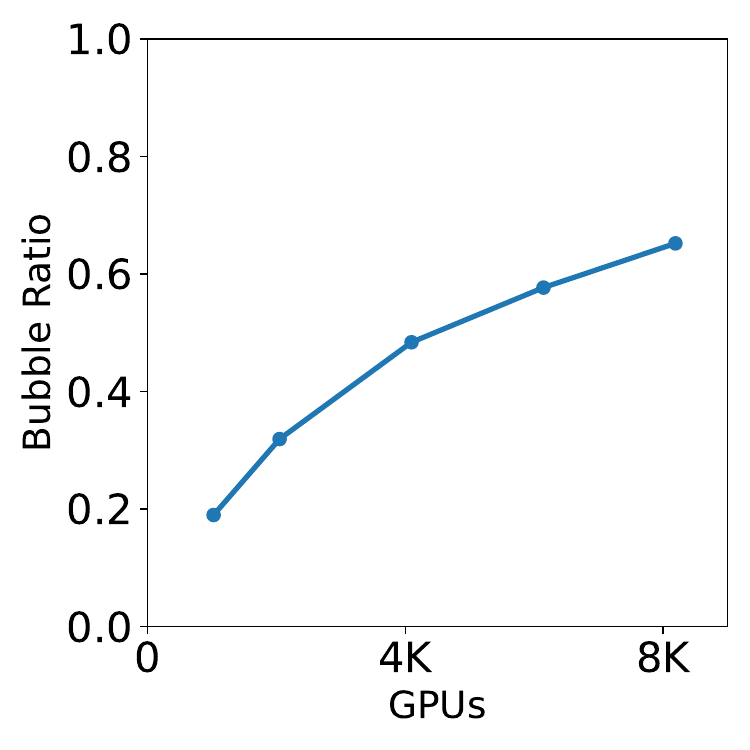}
    \caption{Bubble Ratio v.s. \#~GPUs}
    \label{fig:main_bub}
    \vspace{-10pt}
  \end{subfigure}
  \begin{subfigure}{0.36\linewidth}
    \centering
    \includegraphics[width=\columnwidth]{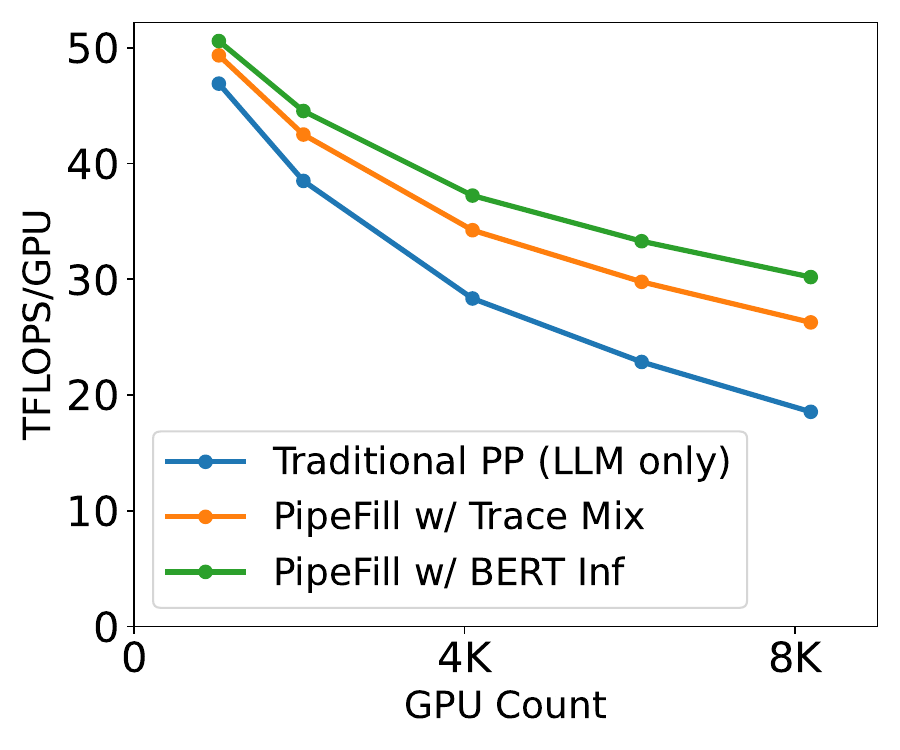}
    \caption{GPU Utilization v.s. \#~GPUs}
    \label{fig:main_res}
    \vspace{-10pt}
  \end{subfigure}
  \caption{Simulator results of running a 40B LLM training job using 1-8K GPUs.}
  \label{fig:main_result}
\end{figure*}

In our evaluation, we first present the amount of GPU utilization recovered by \sysname at different scales (Section~\ref{sec:eval:sim});
we then validate the accuracy of the simulator by comparing simulator results against physical cluster results (Section~\ref{sec:eval:cluster});
we discuss how fill job characterization affects \sysname's performance (Section~\ref{sec:eval:fill});
and we provide sensitive studies of pipeline schedule algorithm, fill-job scheduling policy, bubble duration and free memory (Section~\ref{sec:eval:sen}). 

\subsection{\sysname Recovers GPU Utilization}
\label{sec:eval:recover}

\textbf{Simulator Results}
\label{sec:eval:sim}
To evaluate the GPU utilization recovered by \sysname, we scale the 40B parameter LLM training job trace using data-parallelism up to 8K GPUs in our simulation.
We measure the GPU utilization of filling inference jobs only, and filling both training and inference jobs.
We use the GPU utilization of without using \sysname as the baseline. To calculate the additional GPU FLOPS utilization recovered by \sysname, we use the measured total FLOPs (floating-point operations) executed to complete the fill-jobs (from PyTorch profiling) and divide this by the simulated fill-job completion times (wall-clock time); we average this value across all GPUs across the duration of the main job. \newline
%
Figure \ref{fig:main_result} shows the results of main job training time, pipeline bubble ratio, and GPU utilization from using 1-8K GPUs.
Even at low-scales (1K-2K GPUs), \sysname improves GPU utilization by 5-10\%. However, it is at higher scales that \sysname's potential is shown. Scaling the main job from 2K to 6K GPUs reduces training time from 50 days to 29 days; however, this results in a 40\% drop in GPU utilization. With \sysname, we are able to limit the drop in GPU utilization to $<$23\%. At 4K GPUs (reducing main job training time by 16 days compared to 2K GPUs), \sysname is able to get 89\% of the GPU utilization of traditional pipeline-parallelism at 2K GPUs; at 8K GPUs (reducing main job training time by 9 days compared to 4K GPUs), \sysname is able to get 92\% of the GPU utilization of traditional pipeline parallelism at 4K GPUs. \newline
\sysname's performance is even higher with a more bubble-friendly fill-job workload; in Figure \ref{fig:main_result} we also plot the GPU utilization recovered when filling with only BERT inference jobs. With this workload, \sysname improves GPU utilization by 7.8-15.6\% at low scales (1-2K GPUs). At 4K GPUs, \sysname gets's 96.7\% of the GPU utilization of traditional pipeline-parallelism at 2K GPUs; and at 8K GPUs \sysname exceeds the GPU utilization of traditional pipeline parallelism at 4K GPUs by 6.5\%. These results show that \sysname enables strong-scaling by an additional binary order of magnitude with virtually no loss in GPU utilization, and at higher scales can even increase GPU utilization while strong-scaling. Additionally, due to the high bubble ratios and the relatively modest slowdowns experienced by the fill-jobs, the amount of GPUs worth of work being done by \sysname using only the pipeline bubbles is notable. Depending on the workload, \sysname can run 200-300 GPUs worth of fill-job when when the main job is using 2K GPUs, 600-900 GPUs worth of work when using 4K GPUs, and 1500-2600 GPUs when using 8K GPUs. We discuss this in detail in section \ref{sec:eval:fill}

\greg{agree strongly on needing a caption (which should say enough to make it relatively self-contained (in context) to look at the graph-set and understand what is being shown.  I don't mean discussion of results, but what is evaluated and compared.  Also need (a), (b), (c) for the three graphs with identifying label (even if redundant with text atop the graphs)}
\greg{Note that this (need for caption) is true for all of the graphs.}

\begin{figure}[b]
\centering
\includegraphics[width=7cm,keepaspectratio]{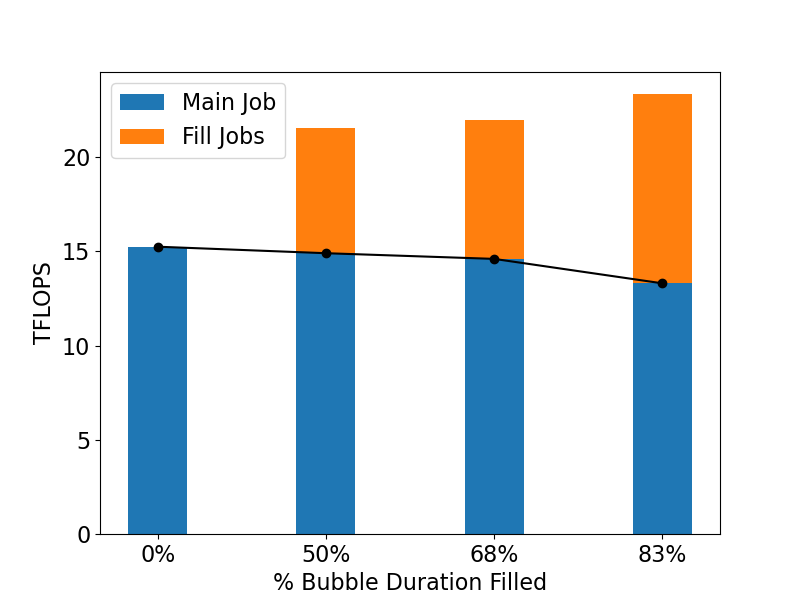}
\caption{GPU TFLOPS of running a 5B LLM on the physical cluster with varying filled bubble durations.}
\label{fig:main_overhead}
\end{figure}

\begin{figure}[h]
\centering
\includegraphics[width=7cm,keepaspectratio]{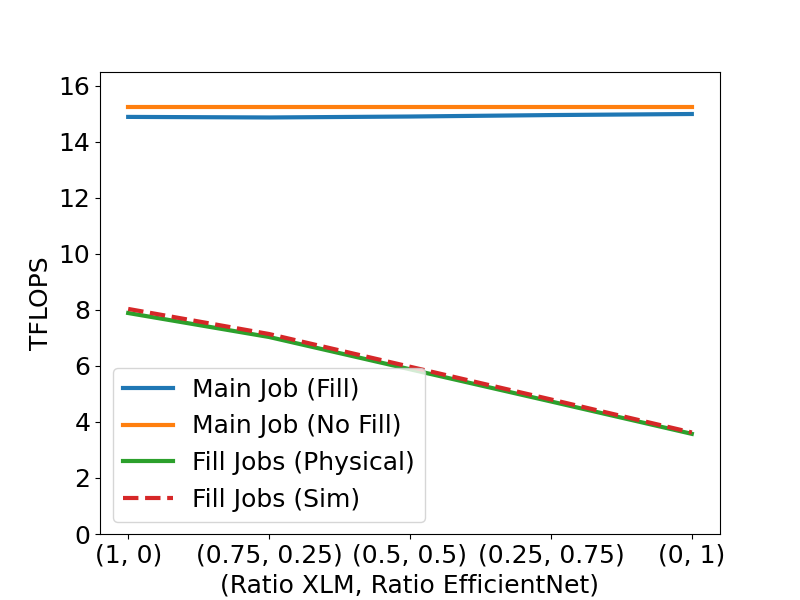}
\caption{Simulator and physical cluster results of running a 5B LLM with varying distributions of fill job types.}
\label{fig:phys_flop}
\end{figure}


\label{sec:eval:cluster}
\noindent \textbf{Physical cluster results.}
We confirm \sysname{}'s effectiveness and validate the fidelity of the simulator results by evaluating a subset of the settings on a small physical cluster with a 5B parameter LLM training job. We measure the free-memory quantity in the bubbles to be 4.5GB without main-job offloading; when we measured the free-memory of the larger training job, it was also 4.5GB, so we use this value in our simulator.
We run the 5B parameter main job using 8~microbatches per minibatch per data-parallel replica; this corresponds to using 64-way data-parallelism and results in a bubble ratio of 65\%, which is comparable to the 8K GPU setting in Figure \ref{fig:main_result}. We also use the full fill-job trace distribution for the physical cluster experiments, unless specified otherwise.

First, we evaluate whether the recovered GPU utilization and low overhead to the main job predicted by the simulator is truly observed in a physical environment. In Figure \ref{fig:main_overhead}, we vary the percentage of the bubble duration that \sysname's Executor's attempt to fill. We find that the overhead to the main job is $<$2\% for up to 68\% of the bubble duration filled by the Executor; at higher fill percentages, the overhead to the main job can be substantial (though the total GPU FLOPS utilization continues to increase). Also at 68\% we see that the TFLOPS/GPU recovered is around 7.39; this is within 5\% of the TFLOPS predicted by the simulator at the same bubble ratio. This is because, in our simulator results, the Fill Job Executors fill the same percentage of the bubble duration by default.

Next, we evaluate whether the types of fill-jobs being run affect the main job overhead. In Figure \ref{fig:phys_flop}, we take two very different job types from our trace: batch-inference with XLM (the largest model) and training with EfficientNet (the smallest model and the only CNN). We fix the percentage of the bubble duration filled by the Executor at 68\%, and vary the fill-jobs from being all XLM to all EfficientNet; we find that the overhead to the main job does not vary significantly. This shows that the overhead to the main job is independent of the types of fill-jobs being executed; instead it is only affected by the percentage of the bubble duration being filled.

Figure \ref{fig:phys_flop} plots the fill-job recovered-FLOPS predicted by the profile-based simulator and observed in physical execution---the maximum error of the simulator is $<$2\%.

\begin{figure*}[t!]
  \centering
  \begin{subfigure}{0.4\linewidth}
    \centering
    \includegraphics[width=\columnwidth]{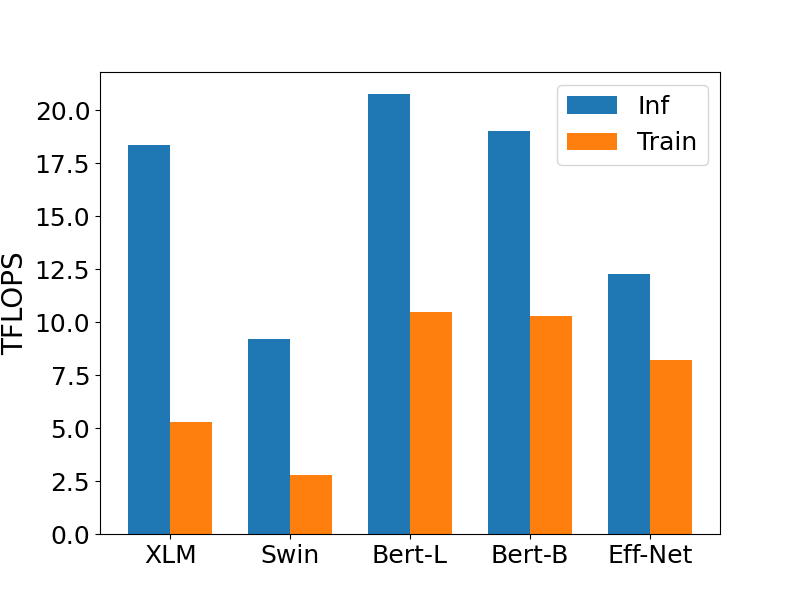}
    \caption{GPU TFLOPS v.s. fill job types}
    \label{fig:fill_flops}
    \vspace{-10pt}
  \end{subfigure}
  \hspace{2mm}
  \begin{subfigure}{0.4\linewidth}
    \centering
    \includegraphics[width=\columnwidth]{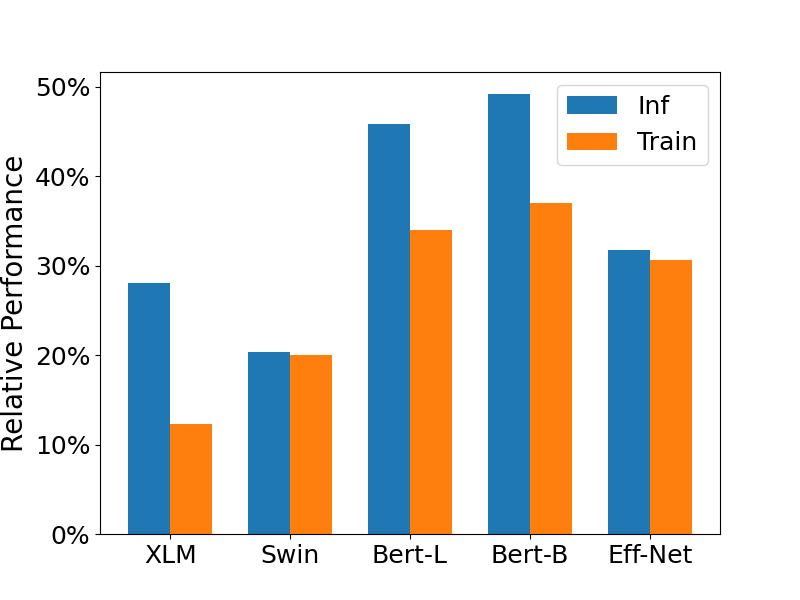}
    \caption{Job slowdown v.s. fill job types}
    \label{fig:fill_slowdown}
    \vspace{-10pt}
  \end{subfigure}
  \caption{GPU TFLOPS and fill job slowdown for different types of fill jobs.}
\end{figure*}

\subsection{Fill job characterization}
\label{sec:eval:fill}

This subsection discusses how fill job characterization affects \sysname's performance.
This study helps understand the tradeoffs in which workloads are used as fill jobs.
In the experiments, we evaluate training and inference of five different models as fill jobs. We measure the GPU FLOPS utilization they are able to achieve during their execution as fill-jobs and compare to the GPU FLOPS utilization they achieve when run in isolated resources. Here we divide the FLOPs (floating-point operations) executed to complete the fill-jobs by the total duration that they are executed (sum of all bubble durations used to complete the fill-job). This is in contrast to dividing by the wall-clock completion time (which we did in section \ref{sec:eval:recover} in order to understand the performance of the fill-jobs when they are executing (as opposed to the FLOPS utilization they can recover). \daiyaan{split this into two sentences, mention as opposed to 6.1}
%

\noindent \textbf{GPU FLOPS.} Different fill jobs are able to utilize the GPU FLOPS to varying degrees; there are several reasons for this, some of which are related to the jobs' fundamental characteristics and some that are related to the bubble constraints. 
In Figure \ref{fig:fill_flops} we plot the GPU FLOPS that each model and each job type (i.e., training vs.\ batch inference) is able to utilize on average during its execution; for comparison, the main job is able to utilize 60 TFLOPS when it is executing. Our first observation is that batch inference jobs are able to reach higher FLOPS utilization than training jobs; this is because inference jobs have low memory requirements and thus can use higher batch sizes under the free memory constraints of the bubbles than training jobs can. Among training jobs, large-model training jobs have particularly poor performance; this is because the much larger activation footprint of these models requires CPU-offloading of the activations. When comparing models, we see that Swin and EfficientNet perform particularly poorly. The Swin model is a non-uniform vision-transfomer model that uses a specialized attention operator; the memory-overhead of the larger layers limit the batch size, which further hurts the GPU utilization of the smaller layers, and the specialized attention operator is not well-optimized in our implementation. 
The EfficientNet model is small compared to the other models, but since it is a CNN it has particularly large activation sizes; the low free-memory in the bubbles limits the batch size that we can use, and since the model is small, the batch sizes that fit in the bubble free-memory are not large enough to reach high GPU utilization.

\noindent \textbf{Fill job slowdown.} TFLOPS recovered lets us compare the GPU utilization recovered across fill-job types, but we would also like to know the slowdown experienced by the fill jobs relative to their performance if they were run on separate, exclusive GPUs. This analysis lets us approximate how many GPUs can be saved during the duration of the main job by filling its bubbles with certain fill-job types.
In Figure \ref{fig:fill_slowdown}, we again see that the slowdown varies substantially across fill-job types. As expected, all fill-jobs experience substantial slowdown due to several factors that put fill-job execution at a disadvantage compared to exclusive execution: 1) the fill-jobs can only use a fraction of the GPU memory (about 25\%) which often necessitates CPU-offloading and limits batch-sizes, 2) the fill-job execution is interrupted every time a bubble ends, introducing unavoidable inefficiencies in the Executor's plan, and 3) because the fill-job execution can only run for a short period of time, each bubble, it often can only run a single iteration of a subset of the model, which is not enough to warmup the GPU caches. \daiyaan{should i mention here that this is why we think we can't fill the whole bubble duration?}\daiyaan{point forward to this from 6.1}However, we see that these factors affect different fill-job types to varying degrees. In particular, we see that although XLM inference recovers similar TFLOPS as BERT inference, it experiences more slowdown; this is because XLM requires aggressive CPU-offloading, but because the model is large it can still submit enough computation work to keep the GPU busy. We hypothesize that on newer hardware-systems that have higher bandwidth between CPU and GPU memory (e.g., newer PCIe generations, NVLink-C2C), the fill-job slowdown from offloading could be substantially lower. Regardless, most of the fill-job workloads we evaluate experience around 30\% of exclusive execution. 

Generally, for a main job using $C$ GPUs with a bubble ratio of $B$ and fill-job relative performance of $P$, we can approximate the GPUs saved by filling as $C*B*P$; for the 8K GPU main job in Figure~\ref{fig:main_result}, \sysname can save over 1500 GPUs for the trace mix and over 2600 GPUs in the best case.

\subsection{Sensitivity studies}
\label{sec:eval:sen}

\begin{figure}[h]
\centering
\includegraphics[width=7cm,keepaspectratio]{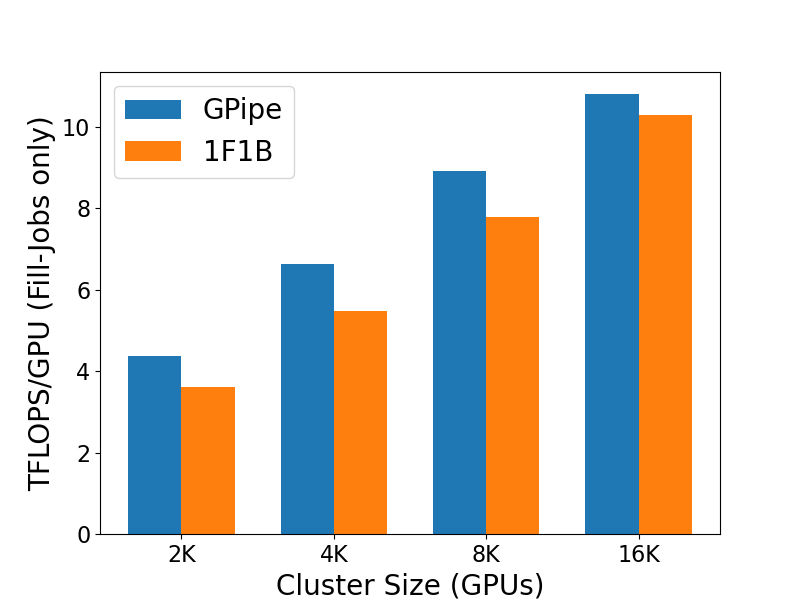}
\caption{
Fill job GPU utilization of using GPipe and 1F1B pipeline schedule algorithms with varying cluster sizes.
}
\label{fig:gpipe_1f1b}
\end{figure}

\noindent \textbf{Main job pipeline schedule.} 
We compare \sysname with the main job using a GPipe schedule to using a 1F1B schedule, using the same main job as the simulator in section \ref{sec:eval:recover} and using the full fill-job trace. We vary the number of GPUs from 2K (18.9\% bubble ratio) to 16K (78.9\% bubble ratio). 
In Figure \ref{fig:gpipe_1f1b}, we see that the at smaller scales \sysname recovers 20\% more GPU utilization when the main job uses GPipe, but at larger scales the gap closes to 5\%. This is because 1F1B contains some non-contiguous bubbles that are not within the fill-drain bubble or the fwd-bwd bubble, which \sysname does not fill; at larger scales these non-contiguous bubbles become a smaller proportion of the total bubbles. \daiyaan{make difference easier to see in figure}

\begin{figure}[h]
  \centering
  \begin{subfigure}{.45\columnwidth}
    \centering
    \includegraphics[width=1\columnwidth]{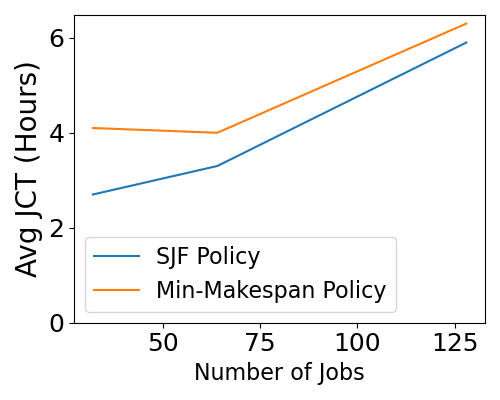}
    \caption{Average JCT}
    \label{fig:placer_jct}
    \vspace{-10pt}
  \end{subfigure}
  \hspace{5mm}
  \begin{subfigure}{0.45\columnwidth}
    \centering
    \includegraphics[width=1\columnwidth]{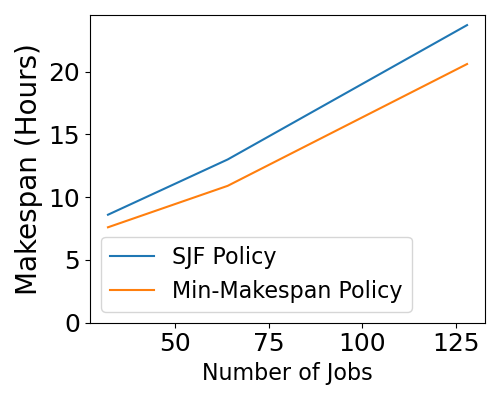}
    \caption{Makespan}
    \label{fig:placer_makespan}
    \vspace{-10pt}
  \end{subfigure}
  \caption{
    Sensitivity study of fill job schedule policy.
    \mason{move the figures towards left a little bit.}
  }
\end{figure}

\noindent \textbf{Fill-job scheduling policy.} 
\sysname allows the scheduling policy to be configured by the user; this section evaluates two possible policies. In Figures \ref{fig:placer_jct} and \ref{fig:placer_makespan}, we implement a Shortest-Job-First policy and a Makespan-Minimizing policy. We see that the SJF policy is able to achieve lower average JCTs, especially at lower loads where completion time is not as dominated by queueing time. Conversely, the Makespan-Minimizing policy is able to reduce makespan, especially at higher loads where maximizing fill-job efficiency has a larger impact.

\noindent \textbf{Bubble durations and free memory.} Main job characteristics affect the pipeline bubble durations and free-memory; for example, a deeper pipeline or a wider main job model (with longer forward and backward times) can increase the bubble durations. Meanwhile, a larger main job model could also reduce the bubble free-memory. Here we analyze the effects of these factors on \sysname's effectiveness.

\begin{figure}[h]
  \centering
  \begin{subfigure}{.45\columnwidth}
    \centering
    \includegraphics[width=1\columnwidth]{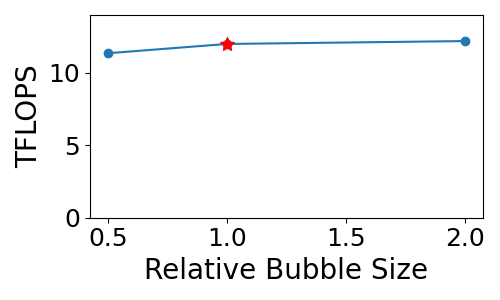}
    \caption{TFLOPS v.s. bubble sizes}
    \label{fig:bub_size}
    \vspace{-10pt}
  \end{subfigure}
  \hspace{2mm}
  \begin{subfigure}{0.45\columnwidth}
    \centering
    \includegraphics[width=1\columnwidth]{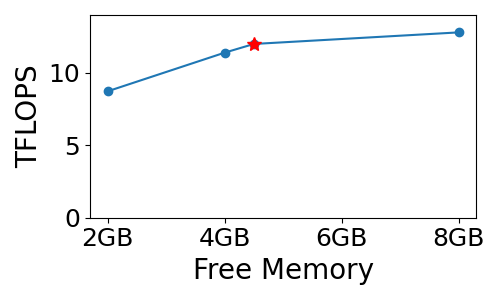}
    \caption{TFLOPS v.s. free mem}
    \label{fig:bub_mem}
    \vspace{-10pt}
  \end{subfigure}
  \caption{
  Sensitivity study of bubble size and free memory.
  \mason{move the figures towards left a little bit.}
  }
\end{figure}

\mason{fix the figure ref}
In Figure \ref{fig:bub_size}, we scale the bubble size by equally scaling the main job model width and depth. We scale the original main job from section \ref{sec:eval:recover}, from 50\% to 200\% of the original model size; we fix the free memory at 4.5GB. We see little difference in the recovered TFLOPS, though shrinking the bubble duration by 50\% reduced TFLOPS by 5.3\%.

\mason{fix the figure ref}
In Figure \ref{fig:bub_mem}, we fix the main job model size (and thus the bubble duration) and vary the free-memory from 2GB to 8GB. We find free-memory to have larger impact on recovered TFLOPS, though with diminishing returns: 4GB recovers 30\% more TFLOPS than 2GB, but 8GB only recovers 12.2\% more TFLOPS than 4GB. 


\section{Related Work}
\label{sec:related}
\noindent \textbf{Pipeline optimizations}.
There are many prior works on increasing pipeline-parallel efficiency. Chimera~\cite{chimera} proposes bidirectional pipelines to reduce pipeline bubbles at the cost of increasing the memory overhead on each device. In practice, it is not possible due to limited GPU memory for large LLM training jobs. Megatron-3D~\cite{megatron} proposes interleaved pipelines, which requires the number of microbatches to be a multiple of the number of pipeline stages. It has limited applicability, since minibatch sizes are fixed, as large-model training is scaled up using data parallelism the number of microbatches per stage decreases quickly to be less than the number of pipeline stages. 
Alpa~\cite{zheng2022alpa}, FlexFlow~\cite{flexflow},  Dapple~\cite{dapple} aim to search for optimal pipeline partition configuration for the training, which cannot eliminate bubbles. Bamboo~\cite{bamboo} introduces redundant computations in pipeline bubbles for distributed training on spot instances. Pipefisher~\cite{pipefisher} uses pipeline bubbles for second-order computations of the training job to accelerate the model convergence. These proposals are orthogonal to our work. 


\noindent \textbf{Resource sharing.}
Many prior works have identified and addressed the data-center GPU under-utilization issue by GPU-sharing. 
AntMan~\cite{antman} provides the elasticity for DL training jobs to scale up and down for better efficiency. Salus~\cite{salus} puts multiple DL jobs on the same device to improve the utilization.
PipeSwitch~\cite{pipeswitch} allows time-sharing of clusters for inference jobs with training jobs, when user demands of inference job is at valley. REEF~\cite{reef} enables kernel-level preemption and concurrent execution for sharing GPUs with multiple inference jobs. PilotFish~\cite{pilotfish} exploits the spare resources on Cloud gaming platform for DL training. Muri~\cite{muri} interleaves the usages of multiple hardware resources (e.g., network, GPU, etc.) among multiple DL jobs. These prior works do not address the pipeline bubbles of large model training like LLMs with tens-of-billions parameters. 

\noindent \textbf{Efficient kernels.} Another common way to improve the compute utilization is to improve the computation efficiency with efficient kernels. FlashAttention~\cite{flashattention, flashattention_v2} improves the computation efficiency and reduces the memory footprints of attention layers by tiling computations. TVM~\cite{tvm}, Ansor~\cite{ansor}, NVFuser~\cite{nvfuser} fuses multiple computations, like elementwise computation with matrix multiplications, to improve the computation occupancy. \sysname{} is orthogonal to this line of research. \sysname{} can take these techniques to further improve the overall utilization for both training and filling jobs. 
\section{Conclusion}
\label{sec:conclusion}

\sysname{} fills the pipeline bubbles of huge DNN training jobs with \emph{other} jobs, significantly reducing the traditional GPU utilization penalty associated with extreme scale-out for such jobs.
%
Experiments confirm that \sysname can increase GPU utilization by up to 63\% when LLM training is scaled-out, with $<$2\% increase in LLM training time, and even gains 5--15\% for low-scale LLM training. 
%
%
Given the explosion of generative AI and the high training costs for the underlying DNNs, \sysname{} provides a critical step forward.

\bibliography{ref.bib}



\end{document}